\documentclass[smallabstract,smallcaptions]{dccpaper}

\usepackage[utf8]{inputenc}
\usepackage[T1]{fontenc}
\usepackage{adjustbox}
\usepackage{epsfig}
\usepackage{amsmath}
\usepackage{amssymb}
\usepackage{color}
\usepackage{url}
\usepackage{listings}
\usepackage{graphicx}
\usepackage{xcolor}
\usepackage{pgfplots}
\usetikzlibrary{shapes,arrows,positioning}
\usepackage{booktabs}
\usepackage{amssymb}
\usepackage[hidelinks]{hyperref}
\usepackage{array}
\usepackage[font={sf},labelfont={sf,bf},justification=centering]{caption}
\usepackage[inline]{enumitem}
\usepackage{verbatim}
\usepackage{wrapfig}
\usepackage{xfrac}
\usepackage{footmisc}

\def\mytitle{Datasets for Benchmarking Floating-Point Compressors}
\def\myauthor{Fabian Knorr, Peter Thoman and Thomas Fahringer}

\hypersetup{
    pdftitle={\mytitle},
    pdfauthor={\myauthor},
    bookmarksnumbered=true,
    bookmarksopen=true,
    bookmarksopenlevel=1,
    colorlinks=false,
    pdfstartview=Fit,
}

\makeatletter
\let\origparagraph\paragraph
\renewcommand\paragraph{\@ifstar{\starparagraph}{\nostarparagraph}}

\newcommand\nostarparagraph[1]
{\paragraphprelude\origparagraph{#1}\paragraphpostlude}

\newcommand\starparagraph[1]
{\paragraphprelude\origparagraph*{#1}\paragraphpostlude}

\newcommand\paragraphprelude{%
  \vspace{-0.5em}
}

\newcommand\paragraphpostlude{}
\makeatother

\begin{document}

\title
{\large
\textbf{\mytitle}
}

\author{%
    \myauthor\\[0.5em]
    {\small\begin{minipage}{\linewidth}\begin{center}
Distributed and Parallel Systems Group\\
        University of Innsbruck, Austria\\
        \url{{fabian,petert,tf}@dps.uibk.ac.at}
    \end{center}\end{minipage}}
}

\maketitle
\vspace{0.5em}
\thispagestyle{empty}

\begin{abstract}
    Compression of floating-point data, both lossy and lossless, is a topic of increasing interest in scientific computing.
    Developing and evaluating suitable compression algorithms requires representative samples of data from real-world applications.
    We present a collection of publicly accessible sources for volume and time series data as well as a list of concrete datasets that form an adequate basis for compressor benchmarking.
\end{abstract}

\vspace*{0.5em}
\Section{Introduction}

Efficient compression of floating-point stream \cite{fpc}\cite{spdp}\cite{mpc} and volume data \cite{lorenzo}\cite{fpzipfull}\cite{ape} has seen significant advances in the past years.
Finding competitive trade-offs between the compression ratios achieved and the computational power invested is the primary objective of this research domain.
Actually benchmarking those quantities relies on suitable input data, which must be similar to the real-world use case of each algorithm. 

Research groups usually rely on their own collection of test data, often without sufficient references to their sources.
This severely limits the comparability of results from different authors, sometimes even rendering precise reproduction impossible.

In this paper, we present a selection of publicly available datasets with appropriate references to aid future publications on the matter.

\vspace*{0.5em}
\Section{Public Data Sources}

\vspace*{0.5em}
\paragraph{Infrared Science Archive} The NASA/IPAC Infrared Science Archive (IRSA)\footnote{\url{https://irsa.ipac.caltech.edu}} is an image data archive for astronomical infrared and submillimeter missions. Among others, it serves images from the Spitzer Space Telescope, which saw prior use in compression research \cite{fpc}.
Data is available in the FITS format \cite{fits}, an image format common in astronomy that supports floating-point data.
The IRSA is funded by the National Aeronautics and Space Administration and operated by the California Institute of Technology.

Spitzer Space Telescope's First Look Survey (FLS) \cite{spitzerfls} and Frontier Fields \cite{spitzerfrontier} data archives contain two-dimensional image files of various sizes in the FITS single-precision floating-point format\footnote{\url{https://irsa.ipac.caltech.edu/data/SPITZER/FLS/images}}\footnote{\url{https://irsa.ipac.caltech.edu/data/SPITZER/Frontier/images}}. 

\paragraph{Radio Telescope Data Center} The Radio Telescope Data Center (RTDC)\footnote{\url{https://www.cfa.harvard.edu/rtdc}} of the Smithsonian Astrophysical Observatory serves data from various US radio telescope sites including the Submillimeter Array and the CfA Millimeter-Wave Telescope.

The Submillimeter Array (SMA) is a radio telescope interferometer for submillimeter wavelength observations located in Hawaii \cite{sma}. 
It is a joint project between the Smithsonian Astrophysical Observatory and the Academia Sinica Institute of Astronomy and Astrophysics and is funded by the Smithsonian Institution and the Academia Sinica.

The CfA Millimeter-Wave Telescope at the Harvard--Smithsonian Center for Astrophysics is an observatory for interstellar molecular clouds \cite{cfa}.

Observational data for both is available as three-dimensional, single-precision FITS floating-point images\footnote{\url{https://www.cfa.harvard.edu/rtdc/SMAimages}}\footnote{\url{https://www.cfa.harvard.edu/rtdc/CO/CompositeSurveys}}.

\paragraph{Hubble Legacy Archive} The Hubble Legacy Archive (HLA)\footnote{\url{http://hla.stsci.edu/}} provides a large selection of imagery from the Hubble space telescope in the FITS image format. Is a joint project of the Space Telescope Science Institute (STScI), the Space Telescope European Coordinating Facility (ST-ECF), and the Canadian Astronomy Data Centre (CADC).

\paragraph{IEEE SciVis Contest} The IEEE Scientific Visualization Contest\footnote{\url{http://sciviscontest.ieeevis.org}} is a yearly installment of the IEEE Visualization Conference (VIS). Simulation datasets from various domains are provided to contestants who submit their own approaches to visualizing them. Some of these datasets are multi-dimensional, binary single-precision floating-point data:
\begin{itemize}[itemsep=0.3em]
    \item SciVis 2004: 3D time steps from simulation of a hurricane from the National Center for Atmospheric Research in the United States\footnote{\url{http://sciviscontest.ieeevis.org/2004/data.html}}.
    \item SciVis 2006: 3D time steps from an earthquake simulation\footnote{\url{http://sciviscontest.ieeevis.org/2006/download.html}}
    \item SciVis 2018: 3D time steps from a of deep water asteroid impact simulation\footnote{\url{https://sciviscontest2018.org/}}.
    \item SciVis 2020: 4D time-step tiles from a simulation of complex eddy transport mechanisms eddies in the Red Sea \cite{scivis2020}\footnote{\url{https://kaust-vislab.github.io/SciVis2020/data.html}}.
\end{itemize}

\paragraph{HDRI Haven} HDRI Haven\footnote{\url{https://hdrihaven.com}} is a source for high-resolution, high-dynamic range (HDR) photographs. Images are available in the Radiance HDR format (also known as RGBE), a shared-exponent floating-point image format. These can be converted to univariate full-width floating point grids by re-mapping the color space, e.g.\ extracting the luminance component. 

\paragraph{Open Scientific Visualization Datasets} Pavol Klacansky provides an online repository of multidimensional simulation data from various domains\footnote{\url{https://klacansky.com/open-scivis-datasets}}. Both single- and double precision datasets exist.

\paragraph{Scientific IEEE 754 Floating-Point Datasets} Martin Burtscher provides test data used in his compression research online. There exist separate pages for single-precision floating-point datasets\footnote{\url{https://userweb.cs.txstate.edu/~burtscher/research/datasets/FPsingle}} used for evaluating the SPDP compressor \cite{spdp} and double-precision floating-point datasets\footnote{\url{https://userweb.cs.txstate.edu/~burtscher/research/datasets/FPdouble}} used, among others, for evaluating the FPC algorithm \cite{fpc}.

\paragraph{UCI Machine Learning Repository}
The UCI Machine Learning Repository is a collection of databases, domain theories and data generators for the empirical analysis of machine learning algorithms. 
\cite{ucimlr} \footnote{\url{https://archive.ics.uci.edu/ml/index.php}}. Some datasets contain time series data, useful for benchmarking one-dimensional compressors.

\paragraph{Freesound} Freesound\footnote{\url{https://freesound.org}} is a collaborative database of audio snippets, samples and recordings. Some data is offered as 32-bit floating-point PCM audio, which corresponds to a single-precision time series.

\paragraph{University of Innsbruck} Simulation data collected at the University of Innsbruck (UIBK) for the purpose of compressor benchmarking can be found on the DPS group website\footnote{\url{https://dps.uibk.ac.at/~fabian/datasets}}.

\Section{Test Datasets}

We propose the collection of data samples from Figure~\ref{fig:datasets} as a baseline for compressor benchmarking. Where no single-precision equivalent exists, double-precision datasets can be truncated to single-precision if the source data range allows it.

\begin{figure}[tb]
\centering \small\renewcommand\arraystretch{0.95}
    \begin{tabular}{>{\ttfamily}lllrr}
        \toprule
        \rmfamily dataset & source & data type & \hspace{-5mm}dimensions & extent \\
        \midrule
        msg\_sppm & Burtscher & single, double & 1 & $34{,}874{,}483$ \\
        msg\_sweep3d & Burtscher & single, double & 1 & $15{,}716{,}403$ \\
        snd\_thunder & Freesound & single & 1 & $7{,}898{,}672$ \\
        ts\_gas & UCI MLR & single & 1 & $4{,}208{,}261$ \\
        ts\_wesad & UCI MLR & single & 1 & $4{,}588{,}553$ \\
        \midrule
        hdr\_night & HDRI Haven & single & 2 & $8{,}192\times16{,}384$ \\
        hdr\_palermo & HDRI Haven & single & 2 & $10{,}268\times20{,}536$ \\
        hubble & HLA & single & 2 & $6{,}036\times6{,}014$ \\
        rsim & UIBK & single, double & 2 & $2{,}048\times11{,}509$ \\
        spitzer\_fls\_irac & IRSA & single & 2 & $6{,}456\times6{,}389$ \\
        spitzer\_fls\_vla & IRSA & single & 2 & $8{,}192\times8{,}192$ \\
        spitzer\_frontier & IRSA & single & 2 & $3{,}874\times2{,}694$ \\
        \midrule
        asteroid & SciVis 2018 & single & 3 & $500\times500\times500$ \\
        astro\_mhd & UIBK & single & 3 & $128\times512\times1024$ \\
        astro\_mhd & UIBK & double & 3 & $130\times514\times1026$ \\
        astro\_pt & UIBK & single, double & 3 & $512\times256\times640$ \\
        flow & Klacansky & double & 3 & $16\times7{,}680\times1{,}0240$ \\
        hurricane & SciVis 2004 & single & 3 & $100\times500\times500$ \\
        magrecon & Klacansky & single & 3 & $512\times512\times512$ \\
        miranda & Klacansky & single & 3 & $1{,}024\times1{,}024\times1{,}024$ \\
        redsea & SciVis 2020 & double & 3 & $50\times500\times500$ \\
        sma\_disk & RTDC & single & 3 & $301\times369\times369$ \\
        turbulence & Klacansky & single & 3 & $256\times256\times256$ \\
        wave & UIBK & single, double & 3 & $512\times512\times512$ \\
        \bottomrule
    \end{tabular}
    \caption{Proposed test datasets, types and grid sizes}
    \label{fig:datasets}
\end{figure}

\begin{itemize}[itemsep=0.3em]
    \item \texttt{msg\_sppm} and \texttt{msg\_sweep3d}, taken from Martin Burtscher's repository, are numeric messages sent by a node in a parallel system running ASCI Purple solvers.

    \item \texttt{snd\_thunder} is a 32-bit float PCM audio recording of thunder, obtained from Freesound user ``Guialgarve''\footnote{\url{https://freesound.org/people/Guialgarve/sounds/523100}}.

    \item \texttt{ts\_gas} is a time series of average temperature-modulated metal oxide gas sensor readings \cite{tsgas} obtained from the UCI Machine Learning Repository\footnote{\url{https://archive.ics.uci.edu/ml/datasets/Gas+sensor+array+under+dynamic+gas+mixtures}}. The file contains readings as truncated floating-point values in text form. The readings from all sensors were averaged to obtain a time series with usable precision.

    \item \texttt{ts\_wesad} is a time series of average readings from physiological and motion sensors during a stress-affect lab study \cite{tswesad}. The data was obtained from the UCI Machine Learning Repository\footnote{\url{https://archive.ics.uci.edu/ml/datasets/Gas+sensor+array+under+dynamic+gas+mixtures}}. The file contains readings as truncated floating-point values in text form. The readings from all sensors were averaged to obtain a time series with usable precision.

    \item \texttt{hdr\_night} and \texttt{hdr\_palermo} are the luminance components of two HDR photographs from HDRI Haven, ``Preller Drive''\footnote{\url{https://hdrihaven.com/hdri/?c=night&h=preller_drive}} and ``Palermo Sidewalk''\footnote{\url{https://hdrihaven.com/hdri/?c=outdoor&h=palermo_sidewalk}}. The images in the largest available resolution were decoded into single-precision floating-point RGB bitmaps and converted to the HSL color space to extract the luminance component.

    \item \texttt{hubble} is an image of the Tadpole Galaxy (UGC10214) from the Hubble Legacy Archive. The dataset id is \texttt{hst\_8992\_03\_acs\_wfc\_f475w}.

    \item \texttt{rsim} is a radiosity field from room response simulation for time-of-flight imaging \cite{rsim}. After extending the implementation to double-precision arithmetic, we simulated the ``Medium'' scene to obtain 2048 time steps.

\item \texttt{spitzer\_fls\_irac}, \texttt{spitzer\_fls\_vla} and \texttt{spitzer\_frontier} are images from the Spitzer First Look Survey and Frontier Fields observations. All datasets were obtained from the Infrared Science Archive.
    \vspace{-0.2em}\begin{itemize}[itemsep=0.3em]
        \item \texttt{spitzer\_fls\_irac} was taken by the Spitzer Infrared Array Camera Component (IRAC) \cite{spitzerirac} as part of the First Look Survey mission. We chose the \texttt{chain1\_main\_mosaic.fits} image.
        \item \texttt{spitzer\_fls\_vla} was taken by the Very Large Array (VLA) radio telescope in preparation of the Spitzer First Look Survey Mission \cite{spitzervla}, for which a single image is available.
        \item \texttt{spitzer\_frontier} is part of the frontier fields observation. Our source is the \texttt{MACS0717.IRAC.1.mosaic.fits} file from the MACS0717 dataset.
    \end{itemize}

\item \texttt{asteroid} is the last time-step of the SciVis 2018 asteroid impact simulation.

\item \texttt{astro\_mhd} is the temperature component of a magnetohydrodynamic simulation of solar wind interactions in the colliding-wind binary system Eta-Carinae \cite{astrocwb}. The simulation was performed separately in single and double precision with a slight variation in size.

\item \texttt{astro\_pt} is one velocity vector component of a particle transport simulation in the LS~5039 system. This simulation was performed separately in single  and double precision as well.

\item \texttt{flow} are the last 16 timesteps of the pressure field of a direct numerical simulation of fully developed flow at different Reynolds numbers in a plane channel \cite{flow}.

\item \texttt{hurricane} is the precipitation component of time step 35 of the SciVis 2004 hurricane simulation.

\item \texttt{magrecon} is a single time step from a computational simulation of magnetic reconnection \cite{magrecon}.

\item \texttt{miranda} is one time step of a density field in a simulation of the mixing transition in Rayleigh-Taylor instability \cite{miranda}.

\item \texttt{redsea} is salt content component from the last time step in the SciVis 2020 contest Red Sea eddy simulation.

\item \texttt{sma\_disk} is observational data of a circumstellar disk from the Submillimeter Array radio inferometer\footnote{\url{https://www.cfa.harvard.edu/rtdc/SMAimages/080403_034904_hd98800.html}}.

\item \texttt{wave} are time steps from a wave propagation simulation on a two-dimensional surface. We modified the \texttt{wave\_sim} simulation code from the Celerity distributed memory runtime \cite{celerity} to support double-precision arithmetic and computed 512 time steps.

\end{itemize}

\Section{References}

\bibliographystyle{IEEEbib}
\bibliography{article}

\end{document}